\documentclass[prl,twocolumn,showpacs,superscriptaddress,preprintnumbers,amssymb]{revtex4}

\usepackage{graphicx}
\usepackage{amsmath}
\usepackage{bm}

\newcommand{\eps}{\epsilon}

\newcommand{\Ref}[1]{Ref.~\cite{#1}}
\newcommand{\Eq}[1]{Eq.~\ref{#1}}
\newcommand{\Fig}[1]{Fig.~\ref{#1}}

\begin{document}

\title{Quantum Phase Transition between the $Z_2$ spin liquid and\\Valence Bond Crystals on a Triangular lattice}

\author{Kevin Slagle}

\author{Cenke Xu}

\affiliation{Department of physics, University of California, Santa Barbara, CA 93106, USA}

\begin{abstract}

We study the quantum phase transition between the $Z_2$ spin
liquid and valence bond solid (VBS) orders on a triangular
lattice. With a fully isotropic triangular lattice, the transition
from a columnar or resonating-plaquette VBS order can be either
first order or there could be two transitions with an intermediate
phase. If the transition splits into two, then the $Z_2$ spin
liquid will first experience a first order $q=3$ Potts transition
to a new nematic $Z_2$ spin liquid that breaks the $2\pi/3$
lattice rotation symmetry (but retain translation symmetry unlike
the VBS states). The second transition will then take this new
nematic $Z_2$ spin liquid to a columnar or resonating-plaquette
VBS state through a second order $3d$ XY$^*$ transition.
On a distorted triangular lattice, the degeneracy between some of
the different columnar VBS orders is lifted, and the phase
transition can reduce to a single $3d$ XY$^*$ transition.

\end{abstract}

\date{\today}

\maketitle

\section{Introduction}

Tremendous progress has been made in the last two years in
searching for spin liquid states in quantum frustrated spin models
by various numerical
methods~\cite{whitekagome,jiang1,jiang2,melko1,melko2}. Using the
topological entanglement entropy~\cite{kitaeventropy,wenentropy},
the $Z_2$ topological liquid phase was identified in the phase
diagram of frustrated spin and quantum boson models on the square
lattice~\cite{jiang1} and Kagome
lattice~\cite{whitekagome,jiang2}. Although the $Z_2$ liquid phase
itself does not break any symmetry, it was confirmed numerically
that a $Z_2$ spin liquid phase can be very close in energy to a
valence bond solid (VBS) state which breaks translation
symmetry~\cite{whitekagome}. Thus it is conceivable that under
weak perturbations the $Z_2$ spin liquid can be driven into a VBS
phase. This liquid-VBS quantum phase transition is what we will
discuss in the current work. In previous works, this liquid-VBS
transition was thoroughly studied on the square
lattice~\cite{sachdevvbs,xubalents}, honeycomb
lattice~\cite{xubalents}, and Kagome lattice~\cite{sachdevkagome}.
The universality class of the liquid-VBS transition in general
depends on the nature of the VBS pattern. On the square and
honeycomb lattice, the quantum phase transitions between a $Z_2$
liquid and simple VBS phases such as the columnar,
resonating-plaquette, and staggered VBS phases have all been
well-understood~\cite{xubalents,xureview}. However, on the
triangular lattice, starting from a $Z_2$ liquid phase, previous
studies only obtained the transition into a $\sqrt{12} \times
\sqrt{12}$ VBS pattern with a large unit
cell~\cite{sondhi1,sondhi2,sondhi3,misguich,Herdman2011}~\footnote{
This VBS pattern has 12 sites in one unit cell on the triangular
lattice~\cite{misguich}. However, in terms of the dual quantum
Ising model on the honeycomb lattice, this pattern has 48 sites
per unit cell~\cite{sondhi2}.}, while a direct transition between
the $Z_2$ liquid and the simple columnar or resonating-plaquette
VBS patterns was never understood in previous theoretical
analysis. This is precisely the gap that we will fill in this paper. 

We choose to study spin systems on the triangular lattice because
the spin-1/2 organic materials with a triangular lattice are the
best experimental candidates for a spin
liquid~\cite{kappa1,kappa2,kappathermo1,kappathermo2,kappafield,131dmit1,131dmit2,131dmit3,131dmitthermo1,131dmitthermo2},
and some of the organic materials in the same family indeed have a
columnar VBS order~\cite{131dmitvbs}. Our analysis predicts that
on an isotropic triangular lattice, the $Z_2$ liquid to the
12-fold degenerate columnar/plaquette VBS order can be either
first order or there could be two transitions with an intermediate
phase (\Fig{meanfield}). The possibility of having either a single
first order transition or two continuous transitions with an
intermediate phase has been observed previously in Fe-pnictide
materials~\cite{dai2008,hunematic,xunematic,qixu,xuhu}. In the
case of two transitions, the $Z_2$ liquid first undergoes a first
order $q=3$ Potts transition to a new nematic $Z_2$ spin
liquid which breaks the $120^{\circ}$ lattice rotation symmetry. A
second transition will then take this nematic $Z_2$ spin liquid to
a columnar or plaquette VBS state through a $3d$ XY$^*$
transition. On a distorted triangular lattice, the VBS pattern
becomes either 2-fold or 4-fold (depending on the details of the
distortion), and the liquid-VBS transition can reduce to one
single 3d XY$^*$ transition. All of these 3d XY$^*$ transitions
should have a very large anomalous dimension of the VBS order
parameter, which can be tested by future numerical simulations.


\section{Model}

Motivated by the recent discovery of the $Z_2$ spin liquid on the
square and Kagome lattices, we expect that the same $Z_2$ spin
liquid can be realized with a certain spin-1/2 Hamiltonian on the
triangular lattice as well. In order to analyze a spin liquid, it
is standard to introduce the slave particles: $\vec{S}_i =
\frac{1}{2} f^\dagger_{i, \alpha} \vec{\sigma}_{\alpha\beta} f_{i,
\beta}$ where $f_{i,\alpha}$ can be either a bosonic or fermionic
spin-1/2 excitation, but either choice is subject to a local
constraint, $\sum_\alpha f^\dagger_{i,\alpha} f_{i,\alpha} = n_i =
1$, in order to match the slave particle Hilbert space with the
spin Hilbert space. This local constraint introduces a continuous
gauge symmetry (U(1) for bosonic spinons and SU(2) for fermionic
spinons), which at low energy can be broken down to a $Z_2$ gauge
symmetry by the mean field state of $f_{i,\alpha}$. The low energy
physics of this state is described by a pure $Z_2$ gauge field. In
this paper we assume that the $Z_2$ spin liquid itself respects
all symmetries of the lattice, which is possible on the triangular
lattice ~\cite{wangashvin}.

In both the $Z_2$ spin liquid and VBS phases, the spin excitation
$f_{i,\alpha}$ is fully gapped. Therefore we can ``integrate out''
the spin excitations and focus on the spin singlet channel of the
system. The spin singlet channel of the system should be described
by the standard $Z_2$ gauge theory on the triangular lattice:
\begin{equation}
H = \sum_{\bigtriangleup, \bigtriangledown} - K \prod_{<ij> \in
\bigtriangleup, \bigtriangledown} \sigma^z_{ij} - \sum_{<ij>} h
\sigma^x_{ij} + \cdots \label{z2hamiltonian}
\end{equation}
where $\sigma^z$ and $\sigma^x$ are ordinary Pauli matrices and
$<ij>$ denotes the {\it link} on the triangular lattice between
sites $i$ and $j$. $\sigma^x=-1$ roughly corresponds to a valence
bond while $\sigma^x=+1$ indicates the absence of a bond.
$\sigma^z$ is therefore the operator that creates/annihilates a
bond. The first term in \Eq{z2hamiltonian} is a ring product of
$\sigma^z$ on every triangle plaquette on the lattice; we will
keep $K > 0$ so that the Hamiltonian (\Eq{z2hamiltonian}) favors
the ring product to be $+1$ on every plaquette. The ellipsis in
\Eq{z2hamiltonian} include all the terms that are allowed by the
$Z_2$ gauge symmetry and lattice symmetry. In particular, we will
add terms that are a product of $\sigma^x$ operators along a
string of neighboring links. The Hamiltonian \Eq{z2hamiltonian} is
invariant under the local $Z_2$ gauge transformation:
$\sigma^z_{ij} \rightarrow \eta_i \sigma^z_{ij} \eta_j$ where
$\eta_i = \pm 1$. The Hilbert space of this system is also subject
to a local constraint, which is analogous to the familiar Gauss
law constraint of an ordinary U(1) gauge field: $\vec{\nabla}
\cdot \vec{E} = \sum_\alpha f^\dagger_{i,\alpha} f_{i,\alpha} =
n_i = 1$. Once the U(1) gauge symmetry is broken down to $Z_2$,
the gauge constraint reduces to
\begin{equation}
\prod_{j = 1 \cdots 6 \text{ around } i} \sigma^x_{ij} =
(-1)^{n_i} = -1
    \label{z2constraint}
  \end{equation}
where $j = 1 \cdots 6$ are the six nearest neighbors of site $i$
on the triangular lattice. Physically, this constraint forces each
site to only share an odd number of bonds. In addition, $h>0$ will
favor only a single bond per site over three or five bonds.
Together, the gauge constraint and $h>0$ therefore roughly
implement a ``hard dimer constraint'' (one site shares exactly one
bond).

\begin{figure}
\includegraphics[width=.35\textwidth]{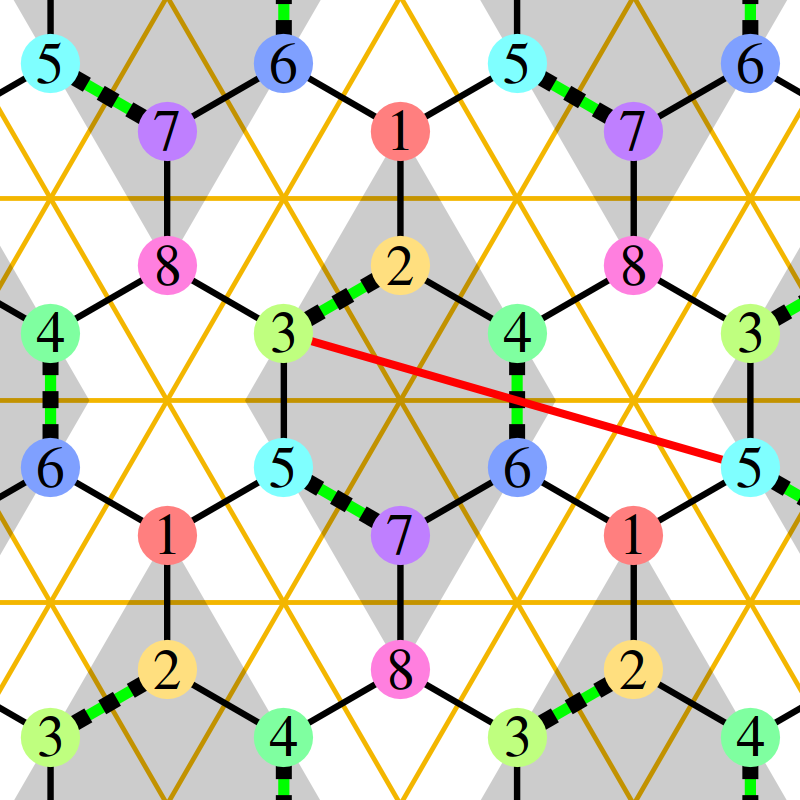}
\caption{ The dual honeycomb lattice on which \Eq{dualhamiltonian}
is defined. $\eta = 1$ on the solid black links, while $\eta = -1$
on the dashed green-black links. Every unit cell (gray diamond)
has eight sites $1 \cdots 8$. In order to obtain the columnar VBS
order, we consider the quantum Ising model defined on this
honeycomb lattice with Ising couplings between sites up to 7th
nearest-neighbor. (One pair of 7th neighbor sites is shown in
red.) }\label{lattice}
\end{figure}

The VBS phase is a confined phase of the slave particle
$f_\alpha$; and in 2+1d $Z_2$ gauge theory, the confinement is
driven by the condensation of vison excitations. A vison is a
plaquette where $\prod \sigma^z = -1$. A vison is not a local
excitation of $\sigma^z$ and $\sigma^x$. Therefore, in order to
describe its dynamics and hence its condensation using a
Landau-Ginzburg-Wilson (LGW) framework, we need to go to the dual
picture where the vison becomes a local excitation. In the dual
language, quantities are defined on the triangular lattice
plaquettes, which form a honeycomb lattice (\Fig{lattice}). The
duality mapping is
\begin{align}
  \tau^x_{q = \bigtriangleup} &= \prod_{<ij> \text{ around } \bigtriangleup} \sigma^z_{ij} \cr\cr
  \sigma^x_{ij} &= \eta_{pq} \tau^z_p\tau^z_q \ \text{ where } p,q \text{ share } <ij>
  \label{duality}
\end{align}
The dual Hamiltonian is a quantum Ising model on the honeycomb lattice:
\begin{equation}
  H = \sum_{<p,q>} - K \tau^x_p - h \eta_{pq} \tau^z_p \tau^z_q + \cdots
  \label{dualhamiltonian}
\end{equation}
Here $p$ and $q$ denote the sites of the dual honeycomb lattice
(plaquettes of the triangular lattice). In this dual
representation, $\tau^a$ are operators while $\eta_{pq} = \pm 1$
are pure numbers. $\tau^x_p$ is the vison density ($\tau^x_p = -1$
means there is one vison at the dual site $p$), and $\tau^z_p$
creates/annihilates a vison on the dual site $p$. The ellipsis in
\Eq{dualhamiltonian} can contain arbitrary further neighbor Ising
couplings allowed by the PSG (described below).

Because of the constraint \Eq{z2constraint}, $\eta_{pq}$ must also
satisfy a constraint: $\prod_{pq \in \mathrm{hexagon}} \eta_{pq} =
-1$, which makes the dual quantum Ising model a fully frustrated
one. Here we choose $\eta_{pq} = -1$ on the dotted links in
\Fig{lattice}, while $\eta_{pq} = +1$ otherwise. The choice of
$\eta_{pq}$ we have made on the dual lattice is just a ``gauge''
choice, which apparently has to break the lattice symmetry; hence
each unit cell on the dual lattice contains eight sites ($1 \cdots
8$). Due to the reduced lattice symmetry in the dual theory, the
correct lattice symmetry transformation for the dual vison
$\tau^z$ must be combined with a nontrivial $Z_2$ gauge
transformation of $\eta_{pq}$ to recover the full symmetry of the
original triangular lattice; this combined transformation is
called the projective symmetry group (PSG)~\cite{wen2002}.
The dual quantum Ising model has to be invariant under the PSG.

The liquid-VBS phase transition corresponds to the Ising
disorder-order phase transition in the dual Hamiltonian
(\Eq{dualhamiltonian}), which is driven by the condensation of
$\tau^z_p$. Because \Eq{dualhamiltonian} is a fully frustrated
quantum Ising model, $\tau^z$ can condense at nonzero momenta in
the dual Brillouin zone (BZ). If \Eq{dualhamiltonian} only has
nearest neighbor hopping (which is the case studied in
\Ref{sondhi2}), then there are four different minimum modes in the
BZ (\Fig{minima}$a$), and the vison condensate corresponds to a
$\sqrt{12}\times \sqrt{12}$ VBS pattern with a large unit
cell~\cite{sondhi1,sondhi2,misguich}. The PSG guarantees that this
liquid-VBS transition belongs to a $3d$ O(4)$^*$ universality
class.

\section{Landau-Ginzburg Analysis}

\begin{figure}
\centering
\includegraphics[width=.33\textwidth]{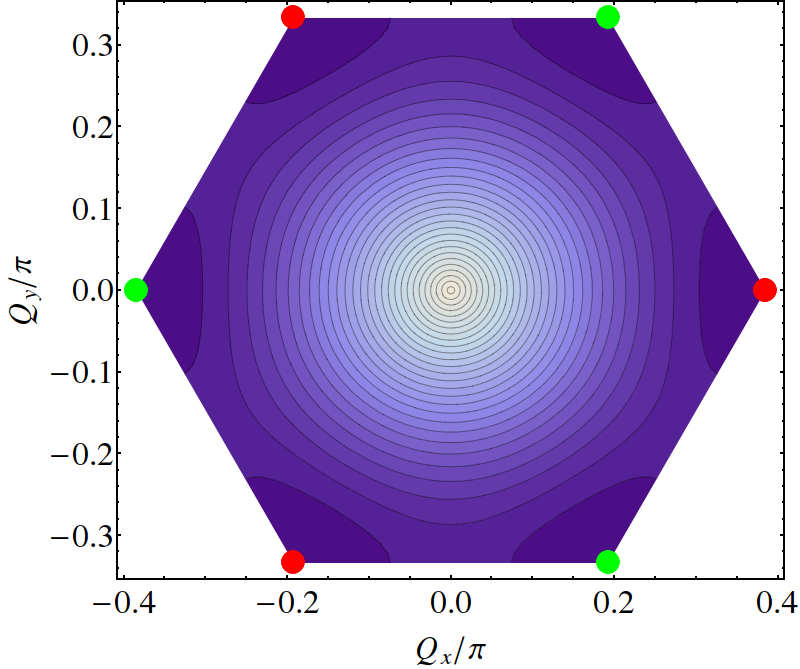} \\ \vspace{1 mm}
{\Large $\quad\;$($a$)} \\ \vspace{3 mm}
\includegraphics[width=.33\textwidth]{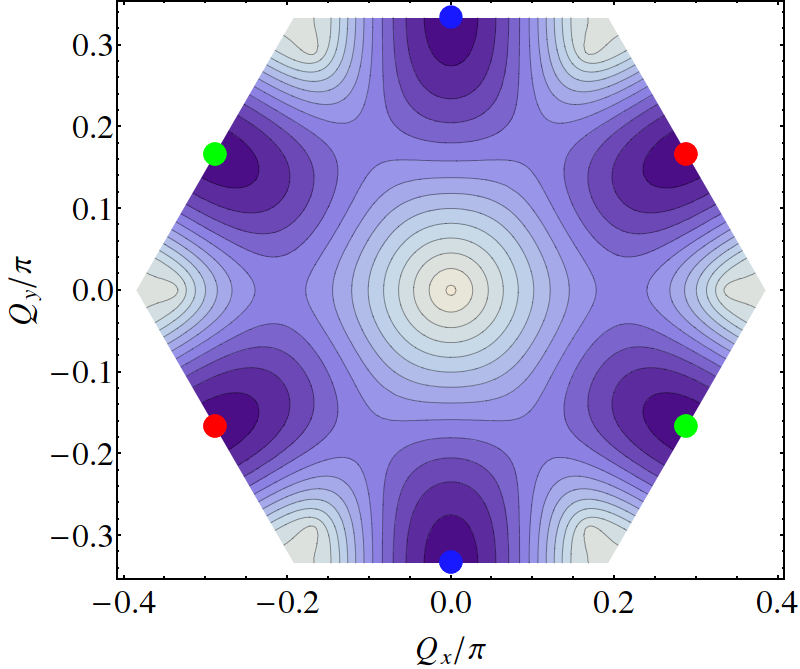} \\ \vspace{1 mm}
{\Large $\quad\;$($b$)} \caption{
The contour plot of the lowest band in
the band structure of the dual quantum Ising model
(\Eq{dualhamiltonian}).
Darker regions correspond to lower energy.
($a$), the band structure with only
nearest neighbor Ising coupling (the case studied in
\Ref{sondhi2}) which has four degenerate minimum modes. ($b$), the
band structure with further neighbor Ising couplings (up to 7th
neighbor) which has six degenerate minimum modes which are stable
against perturbations and weak symmetry reduction. In both ($a$)
and ($b$), each colored dot indicates a degenerate minima with two
different modes (i.e. two orthogonal eigenvectors).
}\label{minima}
\end{figure}

\begin{figure}
\includegraphics[width=3.3 in]{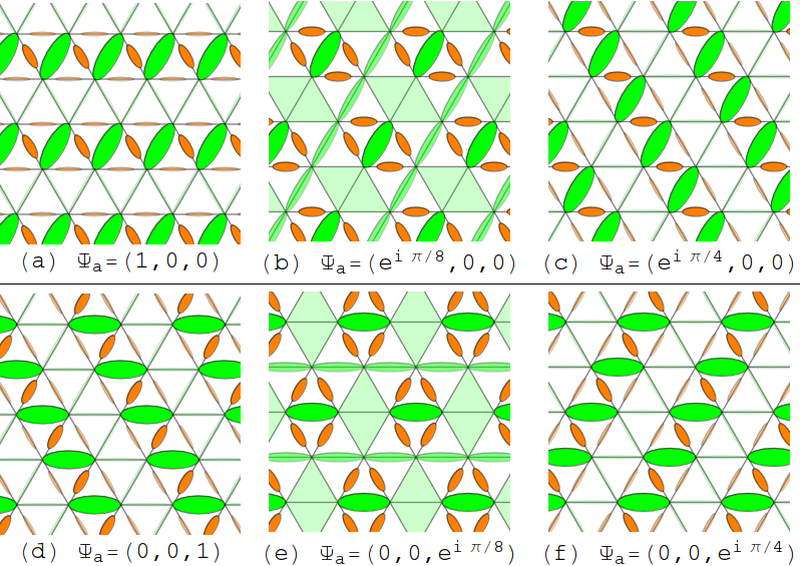}
\caption{ Plot of $\langle \sigma^x \rangle$ on the triangular
lattice based on the ground state of our Lagrangian $\mathcal{L}$
(\Eq{lagrangian}) with $u < - |v|/2$. Green bonds
indicate valence bonds with $\langle \sigma^x \rangle < 0$ while
orange bonds indicate links with $\langle \sigma^x \rangle > 0$.
($a, c, d, f$) show columnar VBS order while ($b, e$) show
resonating-plaquette VBS order. In ($b$) and ($e$), the
resonating-plaquettes are the diamonds highlighted in green.
}\label{vbs}
\end{figure}

Our goal is to study the quantum phase transition between the
$Z_2$ liquid and the simple columnar/plaquette VBS order on the
triangular lattice (\Fig{vbs}). With nearest neighbor Ising
couplings only, the dual Hamiltonian (\Eq{dualhamiltonian}) will
not produce the columnar/plaquette VBS order. We have to turn on
further neighbor couplings in \Eq{dualhamiltonian} that are
allowed by the PSG. We have thoroughly explored the possible
phases of \Eq{dualhamiltonian}. A negative 2nd neighbor Ising
coupling on the dual lattice will destabilize the original order
in \Fig{minima}$a$ at the cost of a ring degeneracy. This ring
degeneracy is not broken until seventh neighbor couplings are
added. With the seventh neighbor couplings, the minima of
\Eq{dualhamiltonian} are then stabilized by six different minimum
modes (\Fig{minima}$b$) with momenta:
\begin{align}
  & \vec{Q}_1 = \vec{Q}_2 = (\frac{\pi}{2\sqrt{3}}, \ -\frac{\pi}{6}) \cr\cr
  & \vec{Q}_3 = \vec{Q}_4 = (\frac{\pi}{2\sqrt{3}}, \ +\frac{\pi}{6}) \cr\cr
  & \vec{Q}_5 = \vec{Q}_6 = (0, \ \frac{\pi}{3})
\end{align}
Each of these momenta correspond to two orthogonal modes; there
are therefore six minimum modes. To analyze the low energy
physics, we expand the Ising operator $\tau^z$ about these six
minimum modes:
\begin{equation}
  \tau^z_{r,n} = \sum_{a=1}^6 \psi_a(r) v_{a,n} e^{i \vec{Q}_a \cdot \vec{r}}
  \label{orderparam}
\end{equation}
$r$ takes on values at the center of the unit cell diamonds
(\Fig{lattice}) and $n = 1 \cdots 8$ denotes the eight sites in
each unit cell. $\psi_a$ with $a = 1 \cdots 6$ are the real fields
corresponding to the six low energy modes and will play the role
of our order parameters. The eight component vectors $v_{a,n}$
(given in the appendix) are the wave functions of $\tau^z_n$ at
each momentum $\vec{Q}_a$. The action of the PSG on $\tau^z_{a,n}$
will induce the action of the PSG on $\psi_a(r)$. $\psi_a$ will
therefore carry a six dimensional representation of the PSG which
will enable us to calculate the lowest order symmetry allowed
Lagrangian.

\begin{figure}
\includegraphics[width=.48\textwidth]{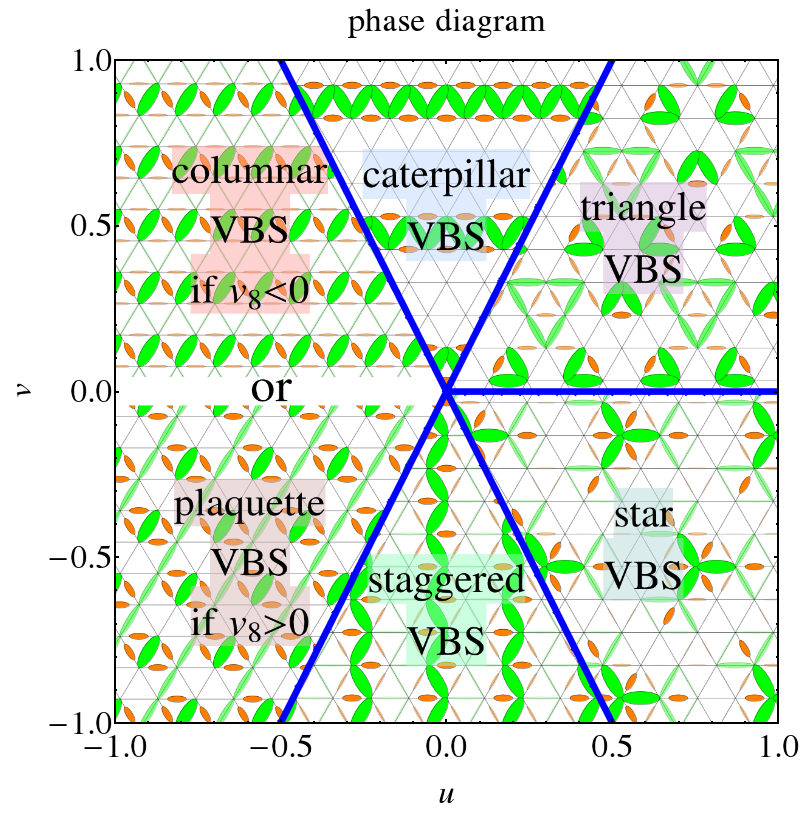}
\caption{ Phase diagram of our Lagrangian $\mathcal{L}$
(\Eq{lagrangian}) in the ordered phase where $r<0$. Green bonds
indicate valence bonds with $\langle \sigma^x \rangle < 0$ while
orange bonds indicate links with $\langle \sigma^x \rangle > 0$. }
\label{phases}
\end{figure}

The PSG is generated by the following transformations (assume the
lattice constant of the original triangular lattice is
$\sqrt{3}$):
\begin{eqnarray}
  T_1       :& x \rightarrow x + \sqrt{3},                                            & \psi_a \rightarrow T_{1,ab} \psi_b \cr\cr
  T_2       :& x \rightarrow x + \frac{\sqrt{3}}{2}, \ y \rightarrow y + \frac{3}{2}, & \psi_a \rightarrow T_{2,ab} \psi_b \cr\cr
  P_x       :& y \rightarrow -y,                                                      & \psi_a \rightarrow P_{ab}   \psi_b \cr\cr
  R_{\pi/3} :& \text{rotation by $\pi/3$}                                             & \psi_a \rightarrow R_{ab}   \psi_b \cr\cr
  TR        :& t \rightarrow -t,                                                      & \psi_a \rightarrow          \psi_a
  \label{psg}
\end{eqnarray}
$R_{\pi/3}$ is a rotation by $\pi/3$ around a hexagon center. The
PSG representation matrices ($T_{1,ab}, T_{2,ab}, P_{ab}, R_{ab}$)
are given in the appendix. The low energy physics of the dual
Hamiltonian (\Eq{dualhamiltonian}) can be completely described by
$\psi_a$ and its effective Lagrangian, which must be invariant
under the PSG. The PSG allowed Lagrangian reads
\begin{align}
  \mathcal{L} &=            \sum_{a=1}^3 \left( |\partial_\mu \Psi_a|^2 + r\, |\Psi_a|^2 \right)
               + g    \left(\sum_{a=1}^3 |\Psi_a|^2 \right)^2 \cr\cr
              &+ u  \,      \sum_{a=1}^3 |\Psi_a|^4
               + v  \,      \sum_{a=1}^3 |\Psi_a|^2 |\Psi_{a+1}|^2 \cos(2\theta_a) \sin(2\theta_{a+1}) \cr\cr
              &+ v_8\,      \sum_{a=1}^3 |\Psi_a|^8 \cos(8\theta_a) + O(\Psi^6)
  \label{lagrangian}
\end{align}
where $\Psi_a = |\Psi_a| e^{i\theta_a}$.
The complex fields $\Psi_a$ are defined as
  $\Psi_1 = \psi_1 + i\psi_2$,
  $\Psi_2 = \psi_3 + i\psi_4$,
  $\Psi_3 = \psi_5 + i\psi_6$ with $\Psi_{a + 3} = \Psi_a$.
The first three terms in $\mathcal{L}$ (\Eq{lagrangian}) are
invariant under an enlarged $O(6)$ rotation of $\psi_a$, while the
$u$, $v$, and $v_8$ terms break this $O(6)$ symmetry down to a
discrete symmetry. There are other 6th and 8th order terms that
are allowed by the PSG, however they are unimportant to both the
critical points and ordered phases considered in this paper. We
always assume that $g$ is the dominant 4th order term in
$\mathcal{L}$ (\Eq{lagrangian}). Under this assumption, there is a
competition between $u$ and $v$, and in the ordered phase ($r<0$)
these coefficients will determine the VBS pattern (see
\Fig{phases} for a phase diagram).

For example, if $u < -|v|/2$, then only one of the three $\langle
\Psi_a \rangle$ will be nonzero. The $v$ term is therefore
irrelevant in this case and the sign of the 8th order $v_8$ term
is necessary to fully determine $\langle \Psi_a \rangle$ in the
ordered phase. The negative and positive $v_8$ correspond to the
columnar and plaquette VBS orders on the triangular lattice
respectively, both of which are 12 fold degenerate. For example,
$\Psi_3 = e^{i \pi n/4}$ and $\Psi_1 = \Psi_2 = 0$ for $n=0,1,2,3$
are four examples of columnar VBS orders. These examples
correspond to four VBS patterns with bonds (links with $\langle
\sigma^x \rangle \sim -1$) aligning in the horizontal direction
(\Fig{vbs}$d$,$f$). Eight more examples are given by taking
$\Psi_1$ or $\Psi_2$ to be nonzero instead of $\Psi_3$. Note that
although taking $n=4,5,6,7$ in $\Psi_3 = e^{i \pi n/4}$ would give
different condensations of $\Psi$, they are actually physically
equivalent to $n=0,1,2,3$. This is because although $\tau^z$ and
$\Psi_a$ differ, $\sigma^x$ is equivalent in these two cases
since the vison fields $\tau^z$ and $\Psi_a$ are only defined up
to an arbitrary $Z_2$ gauge transformation $\Psi_a \rightarrow -
\Psi_a$.

\section{RG Analysis}

Now let us study the nature of the quantum phase transition of
$\mathcal{L}$ (\Eq{lagrangian}). All of the 6th and 8th order
terms will clearly be irrelevant at any critical points and can
therefore be safely ignored. A standard $\eps$-expansion in $d = 4
- \eps$ dimensions of the remaining terms gives us the following
$\beta$-functions:
\begin{align}
  -\beta_r &= 2r - \frac{4}{3}rg - \frac{2}{3}ru \cr\cr 
  -\beta_g &= g\eps - \frac{7}{3}g^2 - \frac{4}{3}gu - \frac{1}{6}v^2 \cr\cr
  -\beta_u &= u\eps - 2gu - \frac{5}{3}u^2 + \frac{1}{12}v^2 \cr\cr
  -\beta_v &= v\eps - 2gv - \frac{2}{3}uv
  \label{betafunc}
\end{align}
To simplify the equations, the above $\beta$-functions and the
table of fixed points below are calculated using rescaled
coefficients as in the following rescaled (and schematic) version
of $\mathcal{L}$ (\Eq{lagrangian}):
\begin{align}
  \mathcal{L}_{\text{RG}}
    &= \frac{1}{2} \left( |\partial\Psi_a|^2 + r\, |\Psi_a|^2 \right) \cr\cr
    &+ \frac{\Omega_d^{-1}}{4!} \left( g\, |\Psi_a^2|^2 + u\, |\Psi_a|^4 + v\, [\cdots] \right)
\end{align}
where 
$\Omega_d = \frac{d\pi^{d/2}}{\Gamma\left(\frac{d}{2}+1\right)} (2\pi)^{-d}$
is the surface area of a $d$ dimensional ball, divided by
$(2\pi)^d$. The above $\beta$-functions have four fixed points
which are given in the following table
\begin{center}
\begin{tabular}{c c c c c}
fixed points    & $r$                 & $g$                & $u$                & $v$ \\[.5ex] \hline
gaussian        & $0$                 & $0$                & $0$                & $0$ \\[1ex]
Ising           & $-\frac{1}{5}\eps$  & $0$                & $\frac{3}{5}\eps$  & $0$ \\[1ex]
Wilson-Fisher   & $-\frac{2}{7}\eps$  & $\frac{3}{7}\eps$  & $0$                & $0$ \\[1ex]
cubic           & $-\frac{3}{11}\eps$ & $\frac{3}{11}\eps$ & $\frac{3}{11}\eps$ & $0$
\end{tabular}
\end{center}
However, expansion of the $\beta-$functions around these fixed
points shows that $v$ is relevant at each of these fixed points.
This implies that none of the $\eps$-expansion fixed points can
harbor a second order phase transition without fine tuning. Thus,
there is likely a run-away flow which suggests that the
transitions described by the dual Lagrangian $\mathcal{L}$
(\Eq{lagrangian}) are first order. \footnote{In principle there is
a chance that higher order $\epsilon-$expansion can lead to a
different result, but we will proceed under the assumption that
the first order result is qualitatively correct.} In the next
section, we will describe why this does not appear to be the
complete story for the columnar or plaquette VBS phase
transitions.

\section{Intermediate Phase}


So far, we have assumed that there is only one phase transition
when a VBS state is driven into a $Z_2$ liquid state. However, it
is possible that there are two transitions with an intermediate
phase in between that breaks less symmetry than the VBS phase.
Without $v$ and $v_8$, $\mathcal{L}$ (\Eq{lagrangian}) has an
emergent $U(1)^3 \times S_3$ symmetry in these phases. The three
copies of $U(1)$ rotate the phases of the three $\Psi_a$ while
$S_3$ will permute the three $\Psi_a$. With a run-away flow under
RG, the $S_3$ and U(1) symmetry can break separately. Since we are
mainly interested in the case with only one of the $\Psi_a$
condenses (the case with negative $u$) which corresponds to the
columnar or plaquette VBS order, there could be an intermediate
nematic phase which only breaks the $S_3$ symmetry. We introduce
an additional complex nematic order parameter $\sigma$ to describe
the $S_3$ symmetry breaking (in addition to the three $\Psi_a$
which will describe $U(1)^3$ symmetry breaking). Thus, the full
Landau-Ginzburg dual Lagrangian is
\begin{align}
  \widetilde{\mathcal{L}}
    &=          \sum_{a=1}^3 \left( |\partial_\mu \Psi_a|^2 + r\, |\Psi_a|^2 \right)
     + g    \left(\sum_{a=1}^3 |\Psi_a|^2 \right)^2 \cr\cr
    &+ u \sum_{a=1}^3 |\Psi_a|^4 + v\, [\cdots] + v_8\, [\cdots] \cr\cr
    &+ |\partial_\mu \sigma|^2 + \tilde{r}\, |\sigma|^2 + \tilde{g}\, |\sigma|^4 - \tilde{u}_3 (\sigma^3 + c.c.) \cr\cr
    &-   \tilde{u}  \left( \sigma^* \sum_{a=1}^3 e^{-2\pi i a/3} |\Psi_a|^2 + c.c.
    \right)+ \cdots
  \label{lagrangiantilde}
\end{align}
with $\tilde{u}_3, \tilde{u}>0$ so that in the ordered phase only one of
$\Psi_a$ condenses.

\begin{figure}
\includegraphics[width=.45\textwidth]{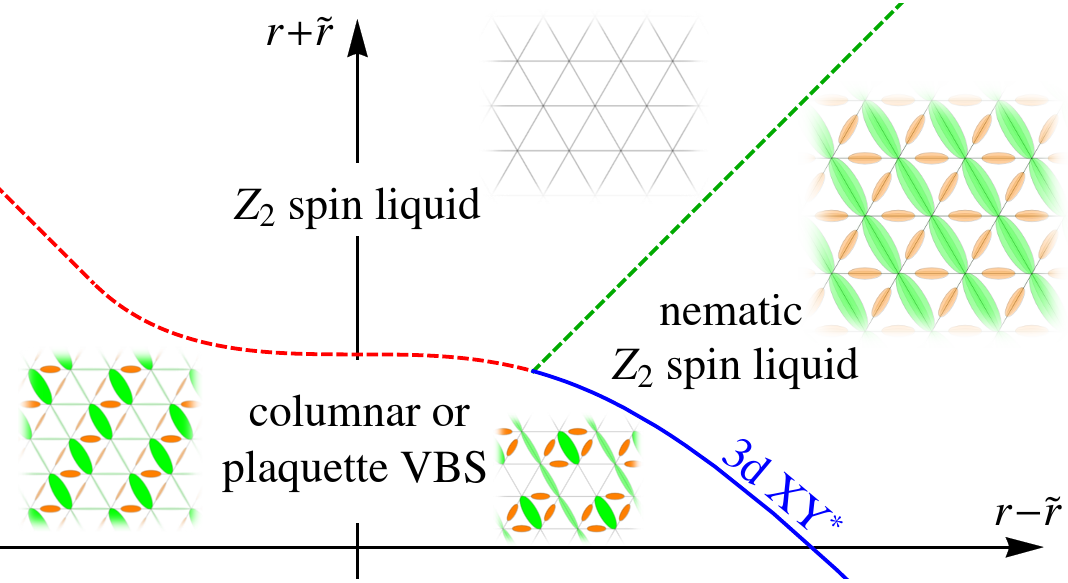}
\caption{ Phase diagram of $\widetilde{\mathcal{L}}$
(\Eq{lagrangiantilde}) and $\widetilde{\mathcal{L}}_{\text{MF}}$
(\Eq{lagrangiantildeMF}) when $u<0$.
Solid lines and dotted lines stand for second and first order
transitions respectively. The nature of the phase transition as $r
+ \tilde{r}$ is varied roughly depends on the sign of $r -
\tilde{r}$. For example, if $r \gg \tilde{r}$, then starting in
the $Z_2$ spin liquid phase, as $r + \tilde{r}$ decreases there
will be a first order transition to an intermediate phase, and
then a second order 3d XY$^*$ transition to a VBS phase.
}\label{meanfield}
\end{figure}

We will now do a mean field analysis of $\widetilde{\mathcal{L}}$
(\Eq{lagrangiantilde}), tentatively neglecting the $v$ and
$v_8$ terms. $\widetilde{\mathcal{L}}$ can then be minimized by
$\Psi_1 = \Psi_2 = 0$ with $\Psi_3 = \psi \ge 0$ and $\sigma \ge
0$ real valued. With this substitution, $\widetilde{\mathcal{L}}$
simplifies to
\begin{equation}
  \widetilde{\mathcal{L}}_{\text{MF}}
     = (r - 2\tilde{u}\, \sigma)\, \psi^2 + \tilde{r}\, \sigma^2
     + (g+u)\, \psi^4 - 2 \tilde{u}_3\, \sigma^3 + \tilde{g}\, \sigma^4
  \label{lagrangiantildeMF}
\end{equation}
The mean field phase diagram of
$\widetilde{\mathcal{L}}_{\text{MF}}$ (\Eq{lagrangiantildeMF}) is
shown in \Fig{meanfield}.

The nature of the phase transition as $r + \tilde{r}$  is varied
roughly depends on the sign of $r - \tilde{r}$. If $r \approx
\tilde{r}$, then $r + \tilde{r}$ will drive the system through a
first order transition from a $Z_2$ spin liquid to a columnar or
plaquette VBS. However, if $r \ll \tilde{r}$ then mean field
predicts a second order transition, while the RG analysis in the
previous section implies that the $v$ term in Eq.~\ref{lagrangian}
will drive this transition first order. On the other hand, if $r
\gg \tilde{r}$ then $\sigma$ will want to order before $\Psi_a$
which will give rise to an intermediate phase. Starting from the
disordered phase ($Z_2$ spin liquid), as $r + \tilde{r}$ decreases
there will be a phase transition to an intermediate ordered phase
with $\langle\sigma\rangle \sim \left\langle \sum_{a=1}^3 e^{-2\pi
i a/3} |\Psi_a|^2 \right\rangle \ne 0$. This phase transition will
spontaneously break the $S_3$ symmetry but keep the $U(1)^3$
symmetry, and the $\tilde{u}_3$ term in
Eq.~\ref{lagrangiantildeMF} will drive the transition to a first
order $q = 3$ Potts transition. Physically, this intermediate
phase will be a nematic $Z_2$ spin
liquid that breaks the $2\pi/3$ lattice rotation symmetry. 
As $r +
\tilde{r}$ is decreased further, there will be another phase
transition which will break the remaining $U(1)$ symmetry. This
phase transition is described by the following theory:
\begin{align}
  \mathcal{L}_3 &= |\partial_\mu \Psi_3|^2 + r\, |\Psi_3|^2 + g\, |\Psi_3|^4 + g_6\, |\Psi_3|^6 \cr\cr
                &+ v_8\, |\Psi_3|^8 \cos(8\theta_3) + O(\Psi^8)
  \label{ani1}
\end{align}
If we view $\Psi_3$ as an order parameter, this transition is in the 3d XY
universality class because $v_8$ is strongly irrelevant.

We have yet to discuss the effects of the topological nature of
the $Z_2$ liquid on any second order phase transitions to a VBS
phase. These effects can be understood by noting that any physical
order parameter must be bilinear of $\psi_a$. This is required
because if $\psi_a \rightarrow -\psi_a$ then $\tau_z \rightarrow
-\tau_z$ (\Eq{orderparam}) but $\sigma^x$ and $\sigma^z$ will
remain invariant. Thus there is a global $Z_2$ gauge redundancy in
the definition of $\psi_a$ and so only quantities that are
quadratic in $\psi_a$ will be gauge invariant. For example, the
columnar and plaquette VBS order parameters are $V_a \sim
\Psi_a^2$. Being quadratic in $\psi$, this order parameter will
have an enormous anomalous dimension $\eta$ where
\begin{equation}
  \langle V_a(x) \ V_a(x') \rangle \sim \frac{1}{|x - x'|^{1 + \eta}}
\end{equation}
Note that $\eta$ must be greater than 1 ($\eta = 1$ in a free
field theory with a bilinear order parameter), which is much
larger than any ordinary Wilson-Fisher fixed point. Therefore, a
3d XY transition in the dual theory is referred to as a 3d XY$^*$
transition in the original theory in order to denote the
difference. Thus, the above 3d XY$^*$ transition should have the
same dynamical exponent $z = 1$ and critical exponent $\nu$ as an
ordinary 3d XY transition, but with a much larger anomalous
dimension $\eta\sim 1.49$ ~\cite{melko2} for the VBS order
parameter.

\section{Anisotropic Triangle Lattices}

All of the previous discussions were under the assumption that the
triangular lattice is fully isotropic. In real materials, a
triangular lattice is usually distorted. For example, in the
triangular lattice spin-1/2 material
Cs$_2$CuCl$_4$~\cite{CsCuCl1,CsCuCl2}, the Heisenberg coupling is
much stronger along one of the three directions. Now let us break
the $\pi/3$ rotation symmetry but keep the translation
($T_1$,$T_2$), reflection $y \rightarrow -y$ ($P_x$), and
inversion $\vec{r} \rightarrow -\vec{r}$ ($R_{\pi/3}{}^3$)
symmetries. This is precisely the symmetry of the material
Cs$_2$CuCl$_4$. The locations of the minima in the BZ are stable
against this symmetry reduction. However, although $\Psi_1$ and
$\Psi_2$ are still degenerate, they are no longer degenerate with
$\Psi_3$; namely the 12 fold degeneracy between different columnar
VBS orders will be lifted.

If $\Psi_3$ is the lowest energy mode ({\it i.e.} the spin
coupling along the horizontal links on the triangular lattice is
stronger than the other two directions), then the PSG allowed
Lagrangian is precisely Eq.~\ref{ani1}. Again, depending on the
sign of $v_8$, the ground state of the VBS phase (the phase with
$r < 0$) is either a columnar (\Fig{vbs}$d$,$f$) or plaquette
(\Fig{vbs}$e$) VBS, both four-fold degenerate. The liquid-VBS
phase transition still belongs to the well-studied 3d XY$^*$
transition.

If $\Psi_1$ and $\Psi_2$ are the lowest energy modes, then the low
energy effective Lagrangian reads
\begin{eqnarray}
  \mathcal{L}_{12}
    &=&         \sum_{a=1,2} \left( |\partial_\mu \Psi_a|^2 + r\, |\Psi_a|^2 \right)
     +  g \left(\sum_{a=1,2} |\Psi_a|^2 \right)^2   \cr\cr
    &+& u \,    \sum_{a=1,2} |\Psi_a|^4
     +  v \,   |\Psi_1|^2 |\Psi_2|^2 \cos(2\theta_1) \sin(2\theta_2)   \cr\cr
    &+& v'      \left( |\Psi_1|^4 \cos(4\theta_1) - |\Psi_2|^4 \cos(4\theta_2) \right)
  \label{p1p2}
\end{eqnarray}
Again we focus on the case with $u < -|v|/2$ when exactly one of
$\langle \Psi_a \rangle$ is nonzero. (This eliminates the role of
the $v$ term in the ordered phase.) In this case, for either
sign of $v'$, the vison condensate is a four-fold degenerate
columnar VBS state (\Fig{vbs}$a$,$c$).

Now let us further reduce the symmetry. For example, if the $P_x$
($y \rightarrow -y$) symmetry is broken while inversion is still
preserved (this is the symmetry of most organic spin liquid
materials), then the columnar VBS order has only a two-fold
degeneracy which only breaks translation symmetry. Now the PSG
allowed Lagrangian reads
\begin{equation}
  \mathcal{L}_3' = |\partial_\mu \Psi_3|^2 + r\, |\Psi_3|^2 + g\, |\Psi_3|^4 + v\, |\Psi_3|^4 \cos(4\theta_3)
  \label{ani2}
\end{equation}
For either sign of $v$, there is a two-fold degenerate columnar
VBS order. In this case the liquid-VBS transition is still the 3d
XY$^*$ transition because it is well-known that the $Z_4$
anisotropy $\cos(4\theta)$ on a 3d XY fixed point
irrelevant~\cite{Oncubic,loubalents}.

Close to the liquid-VBS critical point, since $v_8$ in \Eq{ani1}
and $v$ in \Eq{ani2} are both irrelevant, \Eq{ani1} and
\Eq{ani2} have an emergent $U(1)$ global symmetry. Thus we can
view the VBS order as a superfluid phase that spontaneously breaks
this $U(1)$ symmetry; therefore the liquid-VBS transition can also
be viewed as a liquid-superfluid phase transition. If we approach
this critical point from the superfluid (VBS) side, then this
transition is driven by the proliferation of vortex excitations of
the superfluid phase. In 2+1d space-time, a superfluid phase is
dual to a bosonic QED: a scalar boson (vortex field) coupled to a
2+1d U(1) gauge field. Therefore this liquid-VBS phase transition
is dual to a Higgs transition:
\begin{equation}
\mathcal{L}_{dual} = |(\partial_\mu - i 2 a_\mu)^2\Phi| + r^\prime
|\Phi|^2 + g^\prime |\Phi|^4 + \frac{1}{e^2} f_{\mu\nu}^2
\end{equation}
where $\Phi$ is a complex field that annihilates a pair of
vortices. After condensation ($r^\prime < 0$) $\Phi$ breaks the
$U(1)$ gauge field to a $Z_2$ gauge field. Hence the condensate of
vortex pairs has a $Z_2$ topological order, which is precisely the
topological order of the $Z_2$ spin liquid state we started with.

\section{Conclusion}

In summary, in this work we studied the quantum phase transition
between $Z_2$ liquid and columnar VBS orders on both the isotropic
and distorted triangular lattices. The critical theories proposed
in this work can be checked by future numerical simulations once a
$Z_2$ spin liquid phase is identified on the triangular lattice.
It would also be interesting to study the direct quantum phase
transition from the magnetic order to the columnar VBS orders on
the triangular lattice, which can be viewed as a triangular
lattice generalization of the deconfined quantum critical
point~\cite{deconfinecriticality1,deconfinecriticality2}. This
transition would be driven by condensation of skyrmions or
vortices of the magnetic order. Eventually we also plan to
understand the global phase diagram around the $Z_2$ spin liquid,
which probably involves a noncollinear spiral spin order, the
columnar/plaquette VBS order discussed in this current paper, and
a collinear spin order. A similar global phase diagram was studied
in \Ref{xusachdev} for the case with four vison minima in the BZ,
and we plan to generalize this to our current case with columnar
VBS order. We expect to understand this global phase diagram for
both spin-1/2 and spin-1 systems on the triangular lattice. We
will leave these subjects to future study.

The authors would like to thanks Leon Balents for pointing out
that there could be an intermediate phase between a $Z_2$ spin
liquid and VBS (\Fig{meanfield}). CX is supported by the Alfred P.
Sloan Foundation, the David and Lucile Packard Foundation, the
Hellman Family Foundation, and NSF Grant No. DMR-1151208.

\bibliography{trianglevison}

\appendix{}

\newpage

\section{Appendix}

In this appendix we present more details about the dual frustrated quantum Ising model.
With a 7th neighbor Ising coupling, in a finite region of the phase diagram, the minima of the vison band structure are stabilized at six different minimum modes (\Fig{minima}$b$) with momenta
\begin{align}
  \vec{Q}_1 &= \vec{Q}_2 = (\frac{\pi}{2\sqrt{3}}, \ -\frac{\pi}{6}) \cr\cr
  \vec{Q}_3 &= \vec{Q}_4 = (\frac{\pi}{2\sqrt{3}}, \ +\frac{\pi}{6}) \cr\cr
  \vec{Q}_5 &= \vec{Q}_6 = (0, \ \frac{\pi}{3})
\end{align}
To analyze the low energy physics, we can expand the Ising operator $\tau^z$ at these six minima:
\begin{equation}
  \tau^z_{r,n} = \sum_{a=1}^6 \psi_a(r) v_{a,n} e^{i \vec{Q}_a \cdot \vec{r}}
\end{equation}
where $r$ takes on values at the center of the unit cell diamonds (\Fig{lattice}).
In this equation, $v_{a,n}$ are six eight-component vectors:
\begin{align}
  v_1 &= \left(\begin{matrix} +1 \\ +f \\ +f \\ -f \\ +1 \\ +1 \\ +1 \\ +f \end{matrix}\right), \quad
  v_2  = \left(\begin{matrix} -f \\ -1 \\ +1 \\ +1 \\ +f \\ -f \\ +f \\ +1 \end{matrix}\right), \quad
  v_3  = \left(\begin{matrix} +1 \\ +1 \\ +1 \\ +f \\ +1 \\ +f \\ -f \\ +f \end{matrix}\right)  \cr\cr
  v_4 &= \left(\begin{matrix} -f \\ +f \\ +f \\ +1 \\ -f \\ -1 \\ +1 \\ +1 \end{matrix}\right), \quad
  v_5  = \left(\begin{matrix} +1 \\ -1 \\ +f \\ -1 \\ -f \\ -1 \\ -f \\ +f \end{matrix}\right), \quad
  v_6  = \left(\begin{matrix} -f \\ +f \\ -1 \\ -f \\ +1 \\ -f \\ -1 \\ +1 \end{matrix}\right)
\end{align}
where $f = \sqrt{2}-1$.

The low energy modes $\psi_a$ carry a six dimensional representation of the PSG of the system.
The entire PSG of the system is generated by the transformations in \Eq{psg}.
The representation matrices are
\begin{align}
  T_1 &=
    \left(\begin{matrix}
      -1 & 0 & 0 & 0 & 0 & 0 \\
      0 & +1 & 0 & 0 & 0 & 0 \\
      0 & 0 & 0 & +1 & 0 & 0 \\
      0 & 0 & +1 & 0 & 0 & 0 \\
      0 & 0 & 0 & 0 & 0 & +1 \\
      0 & 0 & 0 & 0 & -1 & 0 \\
    \end{matrix}\right) \cr\cr
  T_2 &=
    \left(\begin{matrix}
      0 & +1 & 0 & 0 & 0 & 0 \\
      -1 & 0 & 0 & 0 & 0 & 0 \\
      0 & 0 & +1 & 0 & 0 & 0 \\
      0 & 0 & 0 & -1 & 0 & 0 \\
      0 & 0 & 0 & 0 & 0 & -1 \\
      0 & 0 & 0 & 0 & -1 & 0 \\
    \end{matrix}\right) \cr\cr
  P_x &=
    \left(\begin{matrix}
      0 & 0 & \frac{1}{\sqrt{2}} & \frac{1}{\sqrt{2}} & 0 & 0 \\
      0 & 0 & \frac{1}{\sqrt{2}} & -\frac{1}{\sqrt{2}} & 0 & 0 \\
      \frac{1}{\sqrt{2}} & \frac{1}{\sqrt{2}} & 0 & 0 & 0 & 0 \\
      \frac{1}{\sqrt{2}} & -\frac{1}{\sqrt{2}} & 0 & 0 & 0 & 0 \\
      0 & 0 & 0 & 0 & \frac{1}{\sqrt{2}} & \frac{1}{\sqrt{2}} \\
      0 & 0 & 0 & 0 & \frac{1}{\sqrt{2}} & -\frac{1}{\sqrt{2}} \\
    \end{matrix}\right) \cr\cr
  R_{\pi/3} &=
    \left(\begin{matrix}
      0 & 0 & 0 & -1 & 0 & 0 \\
      0 & 0 & +1 & 0 & 0 & 0 \\
      0 & 0 & 0 & 0 & 0 & +1 \\
      0 & 0 & 0 & 0 & -1 & 0 \\
      0 & +1 & 0 & 0 & 0 & 0 \\
      -1 & 0 & 0 & 0 & 0 & 0 \\
    \end{matrix}\right)
\end{align}

\end{document}